\documentclass[useAMS,usenatbib]{mn2e}
\usepackage{amssymb}
\usepackage{graphicx}

\title[Polarimetric studies of Comet C/2009 P1 (Garradd)]
{Polarimetric studies of Comet C/2009 P1 (Garradd)}
\author[H. S. Das, B. J. Medhi, S. Wolf, G. Bertrang, P. Deb Roy and  A. Chakraborty ]
{H. S. Das$^{1}$\thanks{E-mail: hsdas@iucaa.ernet.in (HSD)}
, B. J. Medhi$^{2}$ 
, S. Wolf$^{3}$ 
, G. Bertrang$^{3}$
, P. Deb Roy$^{1}$
and A. Chakraborty$^{1}$
\\
$^{1}$ Department of Physics, Assam University, Silchar 788011, India.\\
$^{2}$ Aryabhatta Research Institute of Observational Sciences, Manora Peak, Nainital 263129, India\\
$^{3}$ Institut f\"ur Theoretische Physik und Astrophysik,
Universit\"at zu Kiel, Leibnizstr. 15, 24118 Kiel, Germany
}
\begin{document}

\date{Accepted 2013 September 26. Received 2013 September 23; in original form 2012 December 12}

\pagerange{\pageref{firstpage}--\pageref{lastpage}} \pubyear{2013}

\maketitle

\label{firstpage}

\begin{abstract}
We present the optical  imaging polarimetric observations of comet
C/2009 P1(Garradd) at three different phase angles e.g., 28.2$^\circ$,
28.1$^\circ$  and  21.6$^\circ$.  The  observations  were carried  out
using IUCAA Faint Object Spectrograph and Camera mounted on Cassegrain
focus  of the  2-m Telescope  of IGO,  IUCAA, Pune  in  R$_{comet}$, R
photometric  bands,  on  21 and  22 March 2012  and  ARIES  Imaging
Polarimeter  mounted on  Cassegrain focus  of the  1.04-m Sampurnanand
Telescope of  ARIES, Nainital in R  photometric band, on  23 May 2012.
We  show the presence  of a  jet activity  in the  rotational gradient
treated image of comet  Garradd at phase angle 28.1$^\circ$. These
jets are  mainly oriented  towards the Sun  and extended up  to $\sim$
5100 km from the cometary photocenter. The antisolar extension of
jet seems to be fainter, which is extended up to $\sim$ 1800
km.  It is found that the comet Garradd shows negative polarization at
phase angle  21.6$^\circ$. The  degree of polarization  derived for
Garradd is in good agreement with other comets at nearly similar phase
angles  e.g., comets 67P/Churyumov-Gerasimenko,  22P/Kopff, 1P/Halley,
C/1990 K1  (Levy), 4P/Faye, and  C/1995 O1 (Hale-Bopp) at  phase angle
$\sim$  28$^\circ$,  and 47P/Ashbrook-Jackson  at  phase angle  $\sim$
21.6$^\circ$,  respectively.  It  is  also found  that  the degree  of
polarization of dusty coma of  comet Garradd at phase angle $\sim$
28$^\circ$  is high  but, not  as high  as in  the case  of  comet
Hale-Bopp.

\end{abstract}


\section{Introduction}

The comet C/2009 P1(Garradd) was named after the Australian astronomer
Gordon J. Garradd from Siding  Spring Observatory, Australia, who discovered the
comet on  four images obtained  between August 13.77 and  August 13.81 2009,
at  a heliocentric  distance  of  8.7  AU.  In the same year 2009,
B.  G.  Marsden also calculated the parabolic  orbit
of the comet  Garradd. The aphelion distance of the Garradd is $\sim$ 5000 -- 5500 AU,
which was estimated by Nakano in the year 2011 (Nakano 2011). It  is to  be noted
that the location of inner Oort  cloud is somewhere
in between 2,000  and 20,000 AU. So, the  comet Garradd may belongs  to the inner
Oort cloud family, but it is difficult to comment whether or not comet
Garradd  is dynamically  new.   The comet  reached  its perihelion  on
December 23,  2011 (r  = 1.55 AU) and its closest  approach to the Earth was
on  March 5,  2012 ($\triangle$ = 1.27 AU).
 The visual magnitudes of comet Garradd were 9.7 and 12.4 on March 21 and May 23, 2012 respectively, which was
sufficiently bright enough for observation in optical domain.

The apparition  of the comet Garradd  provided us a very good opportunity  to conduct
different studies using photometric, spectroscopic and  polarimetric techniques to
understand the physical properties of cometary dust grains (Bodewits et  al 2012, DiSanti et al.  2012, Hadamcik et
al. 2012, Kiselev  et al 2012, Tanigawa et al.  2012 and Villanueva et
al. 2012). The preliminary results obtained by the different investigators show
high  production of  dust  in the  comet Garradd, far away from the Sun. All observations indicated that it  belongs to a
\emph{dusty comet  family}. According to Swift's
Ultra-Violet and Optical Telescope's (UVOT) data, at comparable heliocentric distances
the comet Garradd's  behavior is very  similar to the comet  Hale-Bopp (Bodewits et al. 2012).

The polarimetric study of comet over various scattering angles and
wavelengths could provide  very useful information  about the nature of cometary
dust such as composition, albedo, size distribution of grains etc.
(Kikuchi  et al. 1987; Chernova et  al. 1993; Ganesh et al. 1998; Das et al. 2004,
2008a, 2008b, 2010, 2011; Das \& Sen 2006, 2011; Hadamcik et al. 2007, 2010).
In this paper, we present the results of optical polarimetric study of
the comet Garradd at three different scattering angles.
The rest of the paper  is  organized  as  follows.  In  Section   2  we  present
the observations  and  data reduction. Our  results  are presented in section 3.
Finally, a discussion is presented in Section 4, and we conclude with a summary
in Section 5.

\section{Observations and data reduction}

The polarimetric observations of the comet Garradd  were  carried out  at two
different telescopes in India; (1) 2-m Telescope of Girawali Observatory at Inter
University Center for Astronomy and
Astrophysics (IUCAA), Pune (IGO, Latitude: 19$^\circ$5$^\prime$ N,  Longitude: +  73$^\circ$40$^\prime$ E,
Altitude  = 1000 m), and (2) 104-cm Sampurnanand Telescope of  Aryabhatta Research
Institute  of observational sciencES (ARIES), Nainital (AST, Latitude: 29$^\circ$22$^\prime$  N,
Longitude:  79$^\circ$27$^\prime$ E, Altitude  = 1951 m).

\subsection{2-m Telescope, Girawali Observatory, IUCAA}

The 2-m telescope of IGO  has a Cassegrain  focus with a focal  ratio of
$f/10$. The IUCAA Faint Object
Spectrograph and  Camera (IFOSC) was using as a focal plane  instrument for the
polarimetric observation. The imaging was done on
March 21 and 22, 2012 using an  EEV 2K $\times$ 2K pixels
thinned,  back-illuminated  CCD which provides an effective field  of view
of 10.5$^{'}$ $\times$ 10.5$^{'}$ and each pixel corresponds to 0.3  arcsec on the sky.
The gain  and  read out  noise  of the  CCD  are
1.5 e$^{-}$/ADU  and 4 e$^{-}$,  respectively. IFOSC's  capabilities are
enhanced with an imaging  polarimetric mode with a reduced  circular field of view
of  about  2$^{'}$ radius.
It makes  use of a  Wollaston prism and a
half-wave  plate (HWP) in  between camera  lens and  field
lens, which covers the wavelength ranges from  350 to 850 nm.
The  rotatable HWP  gives  components  of  the electric  vector
polarized orthogonally  of varying  intensities after emerging  out of
the Wollaston  prism. The axis  of  Wollaston  prism   is  aligned  to
the north-south (NS) axis of the telescope, and the  HWP is placed in such a
way that  its fast axis aligned  to the axis  of the
Wollaston prism.  Two  successive  positions of  the
HWP  are needed to  obtain the degree of  polarization and
the position  angle of the polarization vector. The optical principle and design of
present IUCAA polarimeter are given by Scarrott et al. (1983) and Sen \& Tandon (1994).
More details on the principles of the instrument are described in Ramaprakash et al. (1998) and Paul et al. (2012).

At a particular angle $\beta$ of the HWP,  the  intensities  of  the  two
orthogonally polarized  beams are measured. When the HWP  rotates an angle
$\beta$,  the electric  vector rotates  through an  angle  2$\beta$. A
ratio R($\beta$) is defined by;

\begin{equation}
R(\beta) = \frac{I_e/I_o - 1}{I_e/I_o + 1} = p~cos(2\theta - 4\beta)
\end{equation}
where $I_e$ and $I_o$ are  extraordinary and ordinary image, $p$ is
the fraction of the total light in the
linearly polarized condition, $\theta$  is the position angle of the plane
of polarization and $\beta$ is the  angle of HWP. When  the  HWP   rotates  successively  22.5$^\circ$,
45$^\circ$ and 67.5$^\circ$ from an initial position 0$^\circ$, the
position  angles  of  the  polarized  components respectively  rotate
45$^\circ$, 90$^\circ$  and 135$^\circ$ from the initial position.
The linear polarization and position angle values could be calculated from  the  normalized Stoke's
vectors $q$ (=  Q/I), $u$ (=  U/I), $q_1$ (= Q$_1$/I) and  $u_1$ (= U$_1$/I)
when the value of $\beta$ is $0^\circ$, $22.5^\circ$, $45^\circ$ and $67.5^\circ$, respectively.

The linear  polarization ($p$) and position angle ($\theta$) of the polarization
vector is defined by;
\begin{equation}
p = \sqrt{(q^2 + u^2)} ~~~~ \textrm{and} ~~~~ \theta = 0.5~\textrm{tan}^{-1}(q/u)
\end{equation}

In principle $p$ and  $\theta$ can be  determined by using  only two
Stoke's vectors $q$ and $u$. But, the additional two values ($q_1$ and  $u_1$) are measured at
$\beta  = 45^\circ$ and  $67.5^\circ$ due  to non-responsivity  of the
system (Ramaprakash et al. 1998).

Both  the narrow-band  filter  (R$_{comet}$,  $\lambda  =  0.684 \ \mu  m$,
$\triangle \lambda = 0.009 \ \mu m$) and broadband filter (R, $\lambda =
0.630 \ \mu  m$, $\triangle  \lambda  =  0.120 \ \mu m$)  were  used for
polarimetric observations.  These filters reduce the contamination by
gaseous species present in the comet.

\subsection{104-cm Sampurnanand Telescope, ARIES}

The 1.04-m Sampurnanand  Telescope of ARIES has a Cassegrain focus with a
focal ratio of  $f/13$. ARIES Imaging Polarimeter (AIMPOL) was using as a back-end
instrument for polarimetric observation, which  consists of a half-wave plate (HWP)
modulator and a beam-splitting Wollaston prism analyzer.
The data were
obtained on May  23, 2012 using  TK  1024$\times$1024 pixels CCD
camera mounted on  the Cassegrain focus of AST.  The observations were
carried out in R filter ($\lambda = 0.630 \ \mu m$, $\triangle \lambda =
0.120 \  \mu m$). Each pixel of the CCD corresponds to 1.73 arcsec and the
field-of-view is  $\sim$ 8  arcmin diameter on  the sky. The  gain and
read  out noise  of the  CCD are  11.98 e$^{-}$/ADU  and  7.0 e$^{-}$,
respectively. The  working principle and design of  AIMPOL are almost similar to IUCAA's imaging polarimeter.
The detail about  AIMPOL and  data  reduction procedures
are available in Rautela et al.(2004), Medhi et al.(2008, 2010) and Pandey et al. (2009).


\begin{table*}
\caption{Log of the observations. Observatory, UT date, heliocentric distance (r), geocentric distance ($\triangle$),
apparent total visual magnitude ($m_v$), phase angle ($\alpha$), position angle of extended Sun -- comet radius
vector ($\phi$), projected diameter for 1 pixel (D), filters and exposure time (time for one exposure $\times$
number of exposures used for the results) during the observations. }
\begin{center}
\begin{tabular}{|c|c|c|c|c|c|c|c|c|c|}
\hline
   Observatory & UT date & r (AU)& $\triangle$ (AU) & $m_v$  & $\alpha$ ($^\circ$)  & $\phi$ ($^\circ$)& D  & Filters & Exposure \\
               &         &       &                  &        &                      &              & (km  px$^{-1}$) &  & time\\
 \hline
IGO, Pune (2 m)   &  March 21, 2012 & 1.96 & 1.36 & 9.71 & 28.2 & 157.7 & 304& R$_{comet}$ & 720s $\times$ 1\\
                  &                 &      &      &      &      &       &    & R           & 180s $\times$ 1\\
IGO, Pune (2 m)   &  March 22, 2012 & 1.97 & 1.38 & 9.75 & 28.1 & 153.9 & 307& R & 60s $\times$ 1, 30s $\times$ 3\\
AST, Nainital (1.04 m) &  May 23, 2012 & 2.52 & 2.75 & 12.41 & 21.6 & 101.1 & 3440& R & 100s $\times$ 10\\
 \hline
\end{tabular}
\end{center}
\end{table*}

%
\begin{table*}
\caption{Results of standard polarized and unpolarized stars at R-filter. p and $\theta$ from literature
(Schmidt et al. 1992; Turnshek et al. 1990 and HPOL). p$_{obs}$ and $\theta$$_{obs}$ from observations.
Offset angle is calculated using the relation: $\theta _0$ = ($\theta$ - $\theta$$_{obs}$).}
\begin{center}
\begin{tabular}{|c|c|c|c|c|c|c|c|}
\hline
   Observatory & UT date & Star & p (\%) & $\theta$ ($^\circ$)   & p$_{obs} (\%)$ & $\theta$$_{obs}$ ($^\circ$) & $\theta _0$ \\
 \hline
IGO   &  March 21, 2012 & HD251204  & 5.33$\pm 0.05$  & 153.7     & 5.30$\pm 0.24$ & 155.8$\pm 1.3$  & --2.1$^\circ$ \\
      &                 & GD319     & 0.09$\pm 0.09$  & 140.0     & 0.37$\pm 0.29$ & 138.6$\pm$22.4 & 1.4$^\circ$\\
      &                 &           &               &              &       &               & \\
      &  March 22, 2012 & HD251204  & 5.33$\pm 0.05$ & 153.7   & 5.30$\pm 0.23 $ & 155.7$\pm 1.24$&  --2.0$^\circ$ \\
      &                 & HD94851    & 0.06$\pm 0.02$     &   NA  & 0.21$\pm 0.09$ & 148.3$\pm$12.5    &  --\\
      \hline
AST   &  May 23, 2012   & HD154445  & 3.96$\pm 0.07$& 99.25 & 3.68$\pm 0.07$ & 88.91$\pm 0.56$ & 10.34$^\circ$\\
      &                 & HD155197  & 4.31$\pm 0.04$&111.59 & 4.27$\pm 0.03$ &102.88$\pm 0.18$ &8.71$^\circ$\\
      &                 & HD154892  & 0.05$\pm 0.03$ & NA   & 0.07$\pm 0.05$ & 22$\pm 20$ & --\\
 \hline
\end{tabular}
\end{center}
\end{table*}


\subsection{Observational procedure}

The geometrical  conditions during the time of  observation are shown
in  the Table 1,  which are  collected from  JPL's HORIZONS  system of
NASA.  The standard stars for null  polarization and  zero-point of
the polarization position angle were taken from Schmidt et al. (1992),
Turnshek et al. (1990) and HPOL
( \verb"http://www.sal.wisc.edu/HPOL/tgts/HD251204.html" ),
respectively.  The results of polarized and unpolarized standard stars
in  R filter  and  their  corresponding values  from  the above  cited
literatures are  depicted in the  Table 2. The degree  of polarization
(p) and position angle ($\theta$) with their corresponding errors from
the literatures  are given in the  fourth and fifth  column, while the
observed  value   of  polarization  (p$_{obs}$)   and  position  angle
($\theta$$_{obs}$)  with their corresponding  errors are  presented in
the sixth  and seventh column of  the Table 2.  The  position angle of the
unpolarized stars HD94851 and HD154891 are not available in the literature and the
polarization value  for HD154891 is available only
for B photometric band.
Since the  zero position  of  HWP  is not  systematically  aligned with  the
north-south direction or with sun-ward direction, the offset angle has
estimated using $\theta _0$  = ($\theta$ - $\theta$$_{obs}$) and given
in eighth column of the Table 2.

By definition the degree of  linear polarization can be either
positive or negative but the value of polarization $p$ obtained
from  the equation  (2)  is  always   positive.  Actually  the   sign  of
polarization  depends   on  the  position   angle  of  the  polarization  plane
with  respect to  the scattering plane in the equatorial reference system.
The  values $P_r$ and  $\theta_r$ for the scattering
plane  could connect  with  the quantities  $P_{obs}$  and  $\theta_{obs}$
using the following relation (Chernova et al. 1993);
\begin{equation}
P_r = P_{obs}. cos 2\theta_r, ~~~~\theta_r = \theta_{obs} - (\phi \pm 90^\circ),
\end{equation}

where $\phi$  is the  position angle of  the scattering plane  and the
sign in the bracket is chosen to ensure the condition $0 \leq \phi \pm
90^\circ  \leq  180^\circ$.  If  $\theta_r$  is  either  $0^\circ$  or
$90^\circ$, the linear polarization  will be either close to the scattering
plane or perpendicular  to it. For AST observation,  the average value
of $\theta_r$ is estimated  to be 129.8$^\circ$. Therefore the polarization
values  will be  negative  in this  case, which is  quite expected  at
$\alpha = 21.6^\circ$.

We have used different astronomical image processing softwares e.g.,
IRAF, IRIS and FITS Viewer to analyze the observed images. Each image
is corrected for bias  and flat.  The photometric center was found with a
precision of ~ 0.1 pixels by fitting the inner part with a Gaussian profile.
The two perpendicularly polarized  images ($I_e$ and
$I_o)$ obtained in a single  plate (for a particular rotation angle of
the  HWP)  are then  trimmed  with  same  dimension (IGO:  176
$\times$ 176  pixels and AST: 41 $\times$ 41  pixels).
Thus, a total  of eight  images having  the  same dimension  are obtained  for
0$^\circ$, 22.5$^\circ$,  45$^\circ$ and 67.5$^\circ$  rotation of the
HWP  for  one set  of  observation.  Since  the  comet  is faint,  the
signal-to-noise  ratio is increased through  building  each polarized
components  by adding  the images  for  each orientation  of the HWP.
In one  series of  observation at R-filter from  IGO on  March 21,  2012
the stellar  trails are visible in the  inner coma region after
the images have been centered and added. So, we have not considered those images
for our analysis.

The   intensity  images   are   obtained  by   adding  two   polarized
components. The total intensity is given by; $I = (I_e + I_o)_{0^\circ}
=  (I_e  +  I_o)_{22.5^\circ}  =  (I_e  +  I_o)_{45^\circ}  =  (I_e  +
I_o)_{67.5^\circ}$.   The  sky  background   is  estimated using standard task in IRAF/IRIS
and the  average  value of background is subtracted from  the scientific image.
The method  of sky subtraction depends  on  the  object  and  sky  condition  at  the  time  of
observation. Background subtraction  is a very difficult task for comets  because the
dusty coma may extends up to the infinity. Usually the sky background is measured in several
positions at the edge of the images, as far as possible, which is about 35000km for IGO and 70000km for AST respectively,
from the photometric center of the comet, keeping in mind that there could be still some signal
of the coma. The sky background is carefully estimated and subtracted for both the cycle of
observations and normalize by transform the images in e$^-$/s.
After normalization, the intensity image  is  treated  by   the  rotational  gradient  technique  to  detect
the image structures.  The  image  emphasizes   the  high  gradient  regions  in
brightness (Hadamcik  et al. 2003a; Hadamcik et  al. 2010). The rotational
gradient creates two  images, one with a radial  shift (in pixel) and
the other  with a  rotational shift (in  degree). The two  images are
added to create  the final image. This method  is useful for enhancing
the low contrast structures that  are radially organized about some points
in  an image.  To study  the high  gradient regions  on  the intensity
images (i.e., to find jet features) treated intensity images have been
built  using Larson-Sekanina's  rotational gradient  technique (Larson
and Sekanina 1984).  Finally, the polarization maps have  been built up
using the images of four polarized components.

\section{Results}
\subsection{Intensity images}
\subsubsection{Radial intensity profiles}

A radial profile  in intensity could  help to  detect the deviation
  from  an isotropic  coma.  To  study the properties of coma  the radial
  surface  brightness profile  is  measured, because  it  is easy  to
  extract  the slopes.  The  intensity of coma could be described by using
  a power law;  $I = k .  r^{q}$  where $k$ is the  scaling factor for
  the coma,  $r$ is the distance  from the photocenter and  $q$ is the
  slope. When $q = - 1$, it is called \textit{canonical slope} (Jewitt
  \& Meech  1987; Goidet-Devel  et al. 1997;  Lamy et al.  2009).  The
  canonical  coma model  of  a nucleus  produces  a steady-state  dust
  outflow that exhibits a constant slope of $-1$.  Actually differences from this value of slope indicate
  the variations from the standard model,
  which  generally  caused by  the temporal  changes  in dust  production,
  radiation  pressure  and  change in  the physical  properties  of the
  grains e.g., sublimation, fragmentation,  changes in  optical properties
  etc.  The  variations of  intensity  versus photocentric distance  in
  log  scale is plotted in Fig. 1(a) and  1(b) for both IGO and
  AST  observations. During  the  entire observation it is noticed that the intensity
  decreases  with  the increasing distance from  photocenter.
  At a small distance from  the photocenter the profiles are dominated
  by  the  atmospheric seeing.   The  seeing  radius  ($r_s$)  for   IGO  and  AST
  observations  were found as  2 arcsec and  4 arcsec, respectively.  The radial
  profile is canonical in nature with  a slope of $-1$ in log  scale
  when the  distance to  the photocenter is,  $(i)$ in between 2000  km to
  25000 km on March 22, 2012 and $(ii)$ in between 8000 km to 30000 km on
  May  23, 2012.   It is to be noted that the canonical  slope can be adjusted by  the subtraction of sky
  background, if necessary.  If the comet  is faint,  then  sometimes the
  signal at the  outer region  of the  coma  may  lesser than the  sky
  background, so care should be taken to subtract the sky background.

\subsubsection{Profiles through the intensity images}
The variation of intensity inside  coma is analyzed by the radial cuts
through  the  intensity images.   The  decrease  in  intensity with  a
gradual  increase in distance from the photocenter is  being well
noticed in both the solar  and antisolar direction (Fig. 1a $\&$ 1b). The intensity
is high in the solar direction as compare to antisolar direction. The changes of slope
for decrease  in intensity  as a  function  of the  photocenter distance in
log  scale is  being  compared for  both  the March  and the  May
observations.
It is found that the slope is dominated by the seeing near the photocenter if the distance is less
than  2000  km  in  IGO  observation  and  less
than 8000  km  in  AST
observation. The slope of the intensity increases gradually with distances from photocenter
in the solar direction whereas slope is found to be decreasing in the antisolar direction for
both the cycle of observations. Our results are in good agreement with Hadamcik et al. (2013).
In March  observations,  the intensity  falls with  an
average slope of $-1$ in the  antisolar direction and $-1.2$ in the solar
direction when  the distance is in between  2000 km to 30000  km from the
photocenter.
But, in the  May observations, the  average slopes
are found to be $-0.95$ when the distance is in between 8000 km to 30000
km in  the antisolar direction  and of about $-1.1$ for the same  range of
distances in the solar direction.


\begin{figure}
\vspace{4cm}
\hspace{1cm}
\includegraphics[width=50mm]{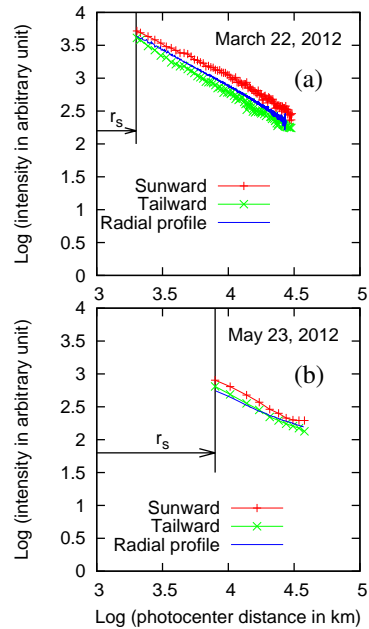}
 \vspace{1cm}
\caption{Cuts through the coma sun-ward, tail-ward and radial profile for the two periods of
observations. Vertical lines represent the seeing radius (r$_s$) limits for each period.
 }
\end{figure}


\subsubsection{Coma morphology}

The  intensity image  with  contours and  their corresponding treated intensity
image using rotational gradient method are shown in Fig. 2a
and 2c, for IGO observation. The position angle
(PA) of the extended Sun and comet radius vector is 153.9$^\circ$.
In the treated image, a strong dust jet feature seems to be
present in the  solar direction with an extension of $\sim$ 5100 km from the
photocenter.  The PA  of the  jet is  about 311$^\circ$.  The tail-ward
extension of the jet is $\sim$ 1800 km and seems to be fainter.

Fig. 2b  and 2d present  the  intensity  image with  contours  and their
corresponding treated image  using rotational gradient  method
for  AST observation. The position
angle  (PA)   of  the  extended   Sun and  comet  radius   vector  is
101.1$^\circ$. The  coma is  extended slightly in  tail-ward direction
and the dust jet feature seems to be absent on May 23, 2012.

\begin{figure}
\vspace{3.5cm}
\hspace{-0.5cm}
\includegraphics[width=80mm]{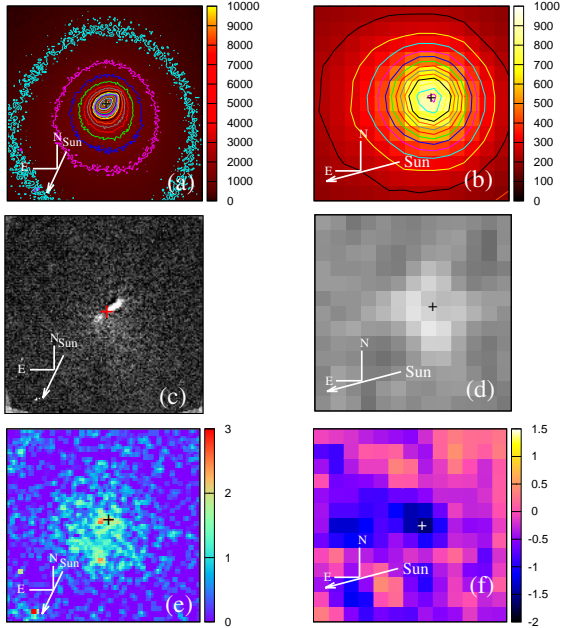}
\caption{(a) and (b): Intensity image with contours (\textbf{the levels are in ADU}) for IGO and AST;
  (c) and (d): The rotational gradient treated image are shown for comet C/2009
  P1 Garradd  for IGO and AST and (e) and (f): Polarization map  (the levels are in \%)  for IGO and AST at phase
  angles 28.1$^\circ$ and 21.6$^\circ$ respectively. The field of view for all the figures is
  45000 km $\times$ 45000 km.  The `+' mark denotes the optical center
  of  the comet.  The position  angles (PA)  of extended  Sun  -- Comet
  radius vector for IGO and AST are 153.9$^\circ$ and 101.1$^\circ$ respectively.  }
\end{figure}

\subsection{Linear polarization}

\subsubsection{Aperture Polarization}

To   study  the   evolution  of  whole-coma   polarization,  the
polarization values  are determined from the  integrated flux obtained
through    all    the     polarized    components    for    increasing
apertures. As discussed earlier, the two perpendicularly polarized images
obtained in a single image are trimmed to the same dimension.
Thus, a  total of eight images having  the same dimension are
obtained  for the four different positions of the HWP. The  eight  images are  properly aligned
, and the magnitude and their corresponding
error at different apertures are computed.
The degree of linear polarization and position angle are then  calculated using
a FORTRAN program based on equations (1) and (2).

We have  only four  sets of  exposures in R-filter  on 22  March 2012.
Since  the  comet is  faint,  the  signal-to-noise  ratio is  enhanced
through building  each polarized components  by adding the  images for
each  orientation  of the  HWP.  Thus the  16  images  are trimmed  to
obtained 16 extraordinary and 16 ordinary images. Finally these images
are  properly  aligned and  combined  to  build eight  perpendicularly
polarized  images. For  the AST  observation,  40 images  are used  to
calculate  the average polarization  value.  The  polarization ($p_{obs}$)
and  the position  angle ($\theta_{obs}$)  obtained  through different
apertures at both the IGO and  AST observations are given in the Table
3. The variation  of polarization in different apertures  is small for
both the narrow-band and broad-band  filters. It is also found that on
21 and 22 March observations,  at phase angle $\sim$ 28$^\circ$ the degree of
polarization  is almost similar  for R$_{comet}$  and R  filter.  This
confirms that the gaseous  contamination by the emission lines through
broadband filter is negligible. For IGO observations, the polarization values obtained from this work
are  in good agreement with Hadamcik et al. (2013).

The polarization values obtained during the observation on May 23 show the negative polarization value  at phase
angle 21.6$^\circ$   as  expected  (already  discussed  in  \emph{Section-2.3}).  In the
Fig. 3, the polarization  values obtained from the present observation
at  $\lambda  =  0.684 \mu  m$  and 0.630 $\mu  m$  are  shown  along with
the polarization values of other comets taken from Kiselev et al. (2006).

\begin{table*}
\caption{Linear Polarization and the position angle ($\theta_{obs}$) of polarization vector for
comet Garradd at different apertures and wavelengths.}
\begin{center}
\begin{tabular}{|c|c|c|c|c|c|c|c|c|}
\hline
   Diameter (in km) & $\rightarrow$  & 4000 & 8000  & 16000 & 20000 & 24000  & 30000& $\theta_{obs}$  \\
   UT Date & Filter  & &   &  &  &  &   &(in degrees)\\
 \hline
 March 21, 2012 & R$_{comet}$  & 2.2$\pm$0.3 & 1.9$\pm$0.2 & 1.9$\pm$0.2 & 2.0$\pm$0.2   & 2.0$\pm$0.3& 2.0$\pm$0.4& 154.1$\pm$4.0\\
 \hline
March 22, 2012  & R  &  2.0$\pm$0.1 & 2.2$\pm$0.1 & 2.3$\pm$0.1 & 2.4$\pm$0.1&2.6$\pm$0.1  & 2.9$\pm$0.2 & 154.6$\pm$1.4\\
 \hline
May 23, 2012 & R  & - & - & -- 0.99$\pm$0.7& -- 0.68$\pm$0.8 & -- 0.95$\pm$0.6 & -- 0.87$\pm$0.6 & 140.8$\pm$9.4\\
 \hline

\end{tabular}
\end{center}
\end{table*}


\begin{figure}
\includegraphics[width=85mm]{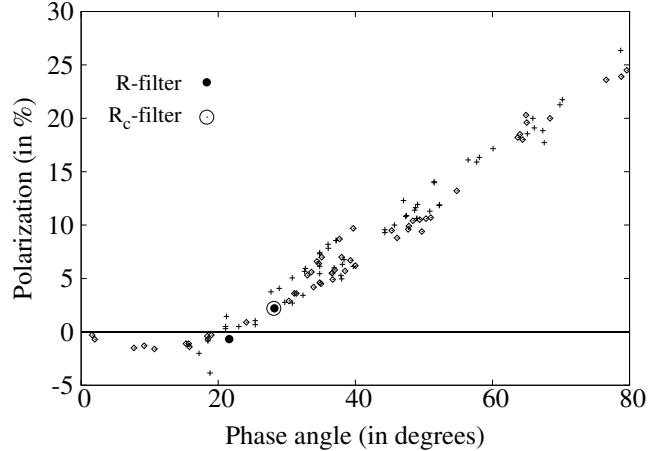}
\caption{Polarization vs phase angle plot for different dusty comets  at  R$_{comet}$-filter
(denoted by `+' symbol) and R-filter (denoted by `$\diamond$' symbol). The symbols `$\bullet$'
and `$\circ$' represent observed polarization value for comet C/2009 P1 Garradd at R-filter
and R$_{comet}$-filter respectively. The linear polarization data of different comets are taken
from Database of Comet Polarimetry by Kiselev et al. (2006).
 }
\end{figure}


\subsubsection{Polarization maps}

The  polarization  maps are created using the four properly aligned polarized  components
of the object.
We built the polarization maps for March 22 and May 23 observations
at  R-filter. It is to be
noted that the polarization map  is obtained using a CL script written
in IRAF and the map is shown
in  Figure 2e  and 2f for IGO  and AST
observations, respectively. The  polarization maps  for both the cycle of observations
are  found to  be in  good  agreement with  the aperture  polarization
values. In March observations the degree of polarization is $\sim$ 3\% at $\alpha =
28^\circ$ when the aperture fits close to photocenter.
But, at  a distance of $\sim$ 4000 km from  the photocenter the degree of
polarization varies from 3\% to 2\%, however  it is
found quite  uniform within $\sim$ 10000 km
from  the  photocenter in  both  solar and  tail-ward
direction. In the outer coma the degree of polarization varies in between
2\% and 0.5\%.  In  the month of May, both the positive and negative polarization  are observed in
the  polarization map  throughout the  coma at $\alpha =
21.6^\circ$.   The degree of polarization
  close to the photocenter is  about $-1.6\%$ and in the
  inner  coma varies  between $-1.8\%$  and $-0.2\%$.  The  positive
  polarization is mainly noticed in  the outer coma and varies between
  0.04 and 1.4\%. The variations of degree of polarization in the inner
  and outer coma suggests that the physical properties of
  cometary dust varies with the distance from the photocenter.

\section{Discussion}

Levasseur-Regourd et al. (1996) studied a polarimetric data-base of 22
comets and  from the  nature of the  phase angle  dependency  concluded
that broadly there are  two  class  of comets exist. But, later on
Hadamcik \&  Levasseur-Regourd (2003b) suggested  that comet Hale-Bopp
itself   represents   an another separate class of comet,  marked   by   unusual   high
polarization. This classification is  prominent only when the phase angles
are higher than 35$^\circ$.  At phase angles between 80 to 100$^\circ$, the
degree of polarization  tends to be  higher than  20\% for a group  of comets
which is known as \emph{Dust rich comets} and smaller than 15\% for another
group which is known  as  \emph{Gas  rich  comets}.
The  observed  linear
polarization values  of the  comet  Garradd at  $\alpha$  = 28$^\circ$  and
21.6$^\circ$  are compared  with  other comets  observed by  different
investigators at similar  phase  angle (Chernova  et al.  1993,
Gural'Chuk et  al. 1987, Kiselev et  al. 2006, Le Borgne  et al. 1987,
Myers 1985, Manset \& Bastein 2000, Renard et al.1996). We found
that the comets 67P/Churyumov--Gerasimenko, 22P/Kopff,  1P/Halley, C/1990
K1 (Levy), 4P/Faye,  and C/1995 O1 (Hale-Bopp) were  observed at phase
angle   close  to   28$^\circ$  and the  comet   47P/Ashbrook-Jackson  at
21.6$^\circ$,  and  our results  are almost  comparable  with  the results
obtained for dusty comets.
The degree of linear
polarization  of  the  dusty  coma   shows  a  high  degree  of  linear
polarization  at phase angle  28$^\circ$,  but  not exceptionally  high  as that of comet
Hale-Bopp.
Kiselev et al. (2012) also made a polarimetric observation of the
comet  Garradd  at different  phase  angles. The degree of polarization obtained
from our data at  $\alpha$
=  21.6$^\circ$ and  28$^\circ$ are in good agreement
with their results at $\alpha$
=  22$^\circ$ and  30$^\circ$, respectively. They also found  that the polarization
of  comet Garradd  measured  in larger  areas  of coma  is almost comparable
with the data available for most of the dusty comets.

The active comets having  well shaped  jets show
higher   polarization   than   less   active   comets   (Hadamcik   et
al.  2003a).  Jets  are  usually  composed  of  very  small  sub-micron size
particles or  large fluffy aggregates. The comets which have higher value of maximum
polarization ($P_{max}$)
show well-structured  silicate features in the 10 $\mu m$ wavelength
domain (Levasseur-Regourd et al. 1996). This suggests the presence of
sub-micron size  particles   or   constituent   grains   in   high   porous
aggregates. The  albedo of these particles is  relatively high (Hanner
1999).  The study  of  comet  Hale-Bopp revealed the existence of  a  highly porous  dust
aggregates  in  the jets  coming  out  from  the nucleus  (Hayward  et
al. 2000, Hadamcik  \& Levasseur-Regourd 2003b), which
are also observed in the  laboratory experiment (Hadamcik et al. 2002)
and  in the  ground  based  observation of  comet  C/1999 S4  (LINEAR)
(Hadamcik  \& Levasseur-Regourd  2003c). In  our observations  of comet
Garradd a sun-ward jet is observed which extends
up to 5100 km in the month of March and also noticed in
the polarization map.

Hadamcik  et  al.  (2012, 2013) has  observed the comet  Garradd  applying  imaging
polarimetric technique and found that
the  polarization  value appear to  be  almost  constant for the  small
apertures  inside   the  coma,   including  the  tail-ward   fan.
Both the
ground-based observations and  space-based observations of comet  Garradd have shown  a high
production of dust far away from the Sun (Villanueva et
al. 2012;  Bodewits et al.  2012).  Bodewits et al. (2012) reported
Swift's  UV-Optical telescope (UVOT)  observations  of  comet Garradd
at  regular intervals at heliocentric  distances between 3.5
and  1.7 AU  on its  inbound trajectory.  From the  UVOT grism
spectra of the comet Garradd they found that its  behavior is nearly similar
to  the  behavior  of   comet  Hale-Bopp  at  comparable  heliocentric
distances.  The polarization  value obtained  from our  observation at
$\alpha = 28^\circ$ is about 2\% for $\lambda = 0.684 \ \mu m$,  but at
such phase angle and wavelength the
polarization  value for the comet Hale-Bopp is $\sim$  4\%
(Manset \& Bastien 2000). It is  clear from our analysis that the
comet Garradd does not show
any  resemblance  with  comet  Hale-Bopp  at such  phase  angle.
This contradicts the result with Bodewits et al. (2012).

 To study the linear polarization properties at small phase angles,
 Hadamcik \& Levasseur-Regourd (2003a)
 built the polarization maps for comets 1P/Wild 2, C/1995 O1 (Hale-Bopp), C/1990 K1
(Levy), 47P/Ashbrook-Jackson and 22P/Kopff.
 In case of  comet 22P/Kopff, it is found that  at  $\alpha =
 18^\circ$  the inner coma  produced a
 strong  negative   polarization  of  $-6\%$  surrounding  the
 photocenter  with  an  extension  of  about   1500  km.  The
 polarization map of comet Hale-Bopp at $\alpha = 19.6^\circ$ shows  a
 negative polarization  of ($-2\pm1$)\% as compared to  the whole coma
 polarization    of   ($1.5\pm0.5$)\%. The negative polarization is
also observed in  comets    C/Levy   and
 47P/Ashbrook-Jackson  near  the inversion  angle i.e.,  at
 $\alpha  =   17.6^\circ$  and  $21.3^\circ$.
Kiselev et al. (2012) also observed the
 negative   polarization   for the comet  Garradd   at phase angle  22$^\circ$   and
 13$^\circ$. They got a minimum  value of about $- 2\%$ near  $\alpha= 13^\circ$.
 Like all those, the negative  polarization has also  been observed for the comet
 Garradd at $\alpha = 21.6^\circ$ in our May observation.
 The polarization map obtained at $\alpha = 21.6^\circ$
 is shown in the Fig. 2f. The innermost coma produces
   negative  polarization of $\sim$  $-1.6\%$, which  is  called as
   \emph{circumnucleus halo},  surrounding a radius of  about 10000 km
   from  the  photocenter.  The positive  polarization  varies
   between 0.04 and  1.4 \%, which is mainly noticed in the outer coma. The
 variation of both  positive and negative polarization  in  all possible direction of coma
 signifies  the  variation  in the  physical properties  of dust in the comet.

\section{Summary}
\begin{enumerate}
\item[(a)] Comet C/2009  P1 (Garradd) has been observed  for three nights in
  the month of  March and May 2012 in polarimetric  mode at the
  phase angles 28.2$^\circ$,  28.1$^\circ$ and 21.6$^\circ$. A small  value  of $\sim$ $- 0.9\%$
polarization  has been  observed  at phase angle
    21.6$^\circ$.

\item[(b)] The radial intensity profile of the comet is canonical
 in nature with a slope of  $-1$ in log scale when the distance to the
 photocenter is $(i)$  in between 2000 km to 25000 km  on March 22, 2012
and $(ii)$ in between 8000 km to  30000 km on May 23, 2013.
The canonical  coma  produces a  steady-state dust outflow.  The decrease in intensity with
a gradual  increase in the photocentric distance is noticed
in  both the  solar and  antisolar  direction of the comet.  On  March 22, 2012  the
intensity  falls  with  an  average  slope of  $-1$ in  the  antisolar
direction and $-1.2$ in the  solar direction when the distance is
in between 2000 km to 30000 km  from the photocenter of the comet. On
May 23, the average slopes are found to be $-0.95$ when the distance is in the
range 8000  km to  30000 km  in the antisolar  direction and  of about
$-1.1$ for the same range of distances in the solar direction.

\item[(c)] The degree of linear polarization of the dusty coma of the comet Garradd
obtained from our observations at R
  and R$_{comet}$ filters are in good agreement with the
  results  reported by  other investigators in similar phase angles.
  The dusty coma shows a high degree
  of  linear polarization  at phase angle 28$^\circ$,  but  not as  high as  the comet
  Hale-Bopp.

 \item[(d)] Jet activity is observed in the treated intensity image of
   the comet Garradd  in the month  of March 2012, which seems to be absent
in the May 2012 observation.  These jets seem  to be
   present in the solar direction with  an extension of $\sim$ 5100 km in the
   projection on the sky. The position angle of the jet is about 311$^\circ$. The
   antisolar   extension of jet  is $\sim$ 1800  km   and  seems   to  be
   fainter.

\end{enumerate}

\section{Acknowledgements}
We  are  highly  grateful   to  anonymous  referee  for  their  valuable
suggestions  and  comments  which  definitely helped  to  improve  the
quality  of  the paper.   We  also  acknowledge  Jeremy R.   Walsh  of
European   Southern   Observatory,   Garching,  Germany   for   useful
discussion.  We gratefully acknowledge IUCAA, Pune and ARIES, Nainital
for making  telescope time  available. This work  is supported  by the
Department of Science \&  Technology (DST), Government of India, under
SERC-Fast Track scheme (FTP/PS-092/2011).

\end{document}